\documentclass[iop,apj]{emulateapj}
\slugcomment{{\sc \underline{Accepted for publication in Geophysical Research Letters:}} October 24, 2015}
\usepackage{amsmath,amssymb,amstext}
\usepackage[breaklinks,colorlinks,citecolor=blue,linkcolor=magenta]{hyperref} 

\usepackage[all]{hypcap} 

\usepackage{natbib}
\shorttitle{Gravity Waves on Mars}
\shortauthors{Kuroda et al.}

\begin{document}

\title{\textbf{A global view of gravity waves in the Martian atmosphere inferred
 from a high-resolution general circulation model}}
\author{Takeshi Kuroda\altaffilmark{1,2}, Alexander S. Medvedev\altaffilmark{2,3},
Erdal Y\.{I}\u{g}\.{I}t \altaffilmark{4,2},
 and Paul Hartogh\altaffilmark{2}}

\altaffiltext{1}{Department of Geophysics, Tohoku University, Sendai, Japan.}

\altaffiltext{2}{Max Planck Institute for Solar System Research, 
G\"ottingen, Germany.}

 \altaffiltext{3}{Institute of Astrophysics, Georg-August University,
G\"ottingen, Germany.}

\altaffiltext{4}{Department of Physics and Astronomy,
 George Mason University, Fairfax, Virginia, USA.}

\begin{abstract}
  Global characteristics of the small-scale gravity wave (GW) field in the Martian
  atmosphere obtained from a high-resolution general circulation model (GCM) are presented
  for the first time. The simulated GW-induced temperature variances are in a good agreement
  with available radio occultation data in the lower atmosphere between 10 and 30 km. The
  model reveals a latitudinal asymmetry with stronger wave generation in the winter
  hemisphere, and two distinctive sources of GWs: mountainous regions and the meandering
  winter polar jet. Orographic GWs are filtered while propagating upward, and the mesosphere
  is primarily dominated by harmonics with faster horizontal phase velocities. Wave fluxes
  are directed mainly against the local wind.  GW dissipation in the upper mesosphere
  generates body forces of tens of m~s$^{-1}$~sol$^{-1}$, which tend to close the simulated
  jets. The results represent a realistic surrogate for missing observations, which can be
  used for constraining GW parameterizations and validating GCM simulations.
\end{abstract}

\keywords{Mars atmosphere, Gravity waves, General Circulation Model, High-resolution modeling}

\section{Introduction}
\label{sec:intro}

The dynamical importance of small-scale gravity waves (GWs) has been well recognized in the
terrestrial middle atmosphere \citep[see the extensive review paper of][]{FrittsAlexander03}
and upper atmosphere \citep[e.g., see the recent review of][]{YigitMedvedev15}. On Mars, GWs
are generated by flow over much rougher than on Earth topography, by strong convection, and
volatile instabilities of weather systems. Amplitudes of Martian GWs are, generally, larger
than those in the lower atmosphere of Earth \citep[e.g.,][]{Cre06a,Wri12} and in the
thermosphere \citep{Cre06b,Fri06}. Upward propagating and ultimately dissipating GWs deposit
a substantial amount of momentum and produce heating and cooling in the Martian middle
atmosphere (50--100 km) and thermosphere (above 100 km) \citep{MY12}. Using the Mars Global
Surveyor (MGS) radio occultation data, \citet{And12} have demonstrated that spectral
amplitudes of small-scale GWs below $\sim$40 km drop off with respect to their vertical
wavenumbers according to the theoretical saturation and power law dependence of $-3$ slope,
which implies a transfer of wave energy and momentum to the mean flow. Based on the MGS
accelerometer data, \citet{Fri06} have found significant body forcing by GWs in the lower
thermosphere.

The Martian atmosphere is approximately 100 times less dense than the terrestrial
one. Accordingly, molecular viscosity is to the same degree larger on Mars, and damping by
molecular diffusion and thermal conduction must be taken into account when GW propagation is
considered, as in Earth's thermosphere. GWs of interest have horizontal wavelengths usually
smaller than the conventional resolution of general circulation models (GCMs), and, thus,
their effects have to be parameterized. \citet{Medvedev_etal11a} applied the nonlinear
spectral parameterization of small-scale GWs of \citet{Yigit_etal08} to the output of the
Mars Climate Database \citep{GG09} and demonstrated that dynamical effects of these waves in
the Martian lower thermosphere are very large and, therefore, cannot be ignored. This
parameterization was specifically developed for ``whole atmosphere" GCMs, and was
extensively utilized in numerous GW studies in the context of Earth's middle atmosphere and
thermosphere \citep{Yigit_etal09,Yigit_etal12,Yigit_etal14,YigitMedvedev09,
  YigitMedvedev12}. With the parameterization interactively implemented into the Max Planck
Institute Martian GCM (MGCM) \citep{Har05, Hartogh_etal07, MH07}, \citet{Med11b} have shown
that GWs play a very important role in the dynamics of the middle and upper atmosphere of
Mars. They close, and even reverse, the zonal jets, enhance the meridional circulation and
middle atmosphere polar warmings, facilitate a formation of CO$_2$ ice clouds
\citep{Yigit_etal15}, and modulate the upper atmospheric response to dust storms
\citep{Medvedev_etal13}. GW-induced cooling is as strong in the mesosphere and thermosphere
as the major radiative cooling mechanism -- the radiative transfer in the IR bands of CO$_2$
molecules \citep{Medvedev_etal15}, -- and can explain the observed temperatures in the lower
thermosphere \citep{MY12}.

GW parameterizations assume a spectrum of wave harmonics at a certain source level in the
lower atmosphere in order to represent GW generation and activity. Accurate estimates of GW
momentum fluxes have, therefore, been recognized as an essential task in the Earth climate
studies. However, with the concerted efforts and numerous observational campaigns
\citep{Alexander10}, the global picture of GWs is still beyond our reach even on Earth. On
Mars, this goal is even farther away. The progress with numerical modeling has allowed to
circumvent this problem to a certain degree by utilizing high-resolution (GW-resolving)
GCMs. They are now being increasingly used in Earth studies for the interpretation and
validation of observations and constraining parameterizations
\citep[e.g.,][]{Watanabe08,Sat09,Miyoshi14}. This approach is based on the assumption that
comprehensive GCMs can capture a significant portion of GW sources and the details of wave
propagation. Thus, they provide a realistic surrogate for observations.

The first high-resolution GCM for Mars has been reported by \citet{Tak08}, however, GWs have
not been considered explicitly at that time.  The only other high-resolution MGCM has been
presented by \citet{Miy11}.  They performed simulations with a horizontal resolution of
$2\times 2$ degrees, and analyzed spatio-temporal spectra of the resolved fields.  The major
finding of their work was an enhancement of wave energy for harmonics with zonal wavenumbers
$s$ up to 30 at tidal frequencies at heights where diurnal and semidiurnal tides are large.
Our paper further addresses the lack of knowledge of GW fields in the Martian atmosphere
with the new high-resolution ($\sim$1.1 degrees in horizontal) MGCM, and directly focuses on
smaller-scale ($s>60$) harmonics, which usually have to be parameterized.

The paper is structured as follows. The high-resolution MGCM is described in
Section~\ref{sec:model}. GW variations in the lower atmosphere (10--30 km) are presented and
compared with observations in Section~\ref{sec:LowerAtm}. Vertical propagation of GWs is
discussed in Section~\ref{sec:vert}, while horizontal distributions of their characteristics
are given in Section~\ref{sec:horiz}.

\section{Gravity-Wave Resolving Martian General Circulation Model}
\label{sec:model}

The high-resolution MGCM used in this study is based on the atmospheric
component of the MIROC (Model for Interdisciplinary Research On Climate)
terrestrial GCM developed collaboratively by the Atmosphere and Ocean Research
Institute (AORI), The University of Tokyo, the National Institute of
Environmental Studies (NIES), and the Japan Agency for Marine-Earth Science 
and Technology (JAMSTEC) in Japan \citep{K-104,Sak12}. It utilizes a spectral 
solver for the three-dimensional primitive equations, and has a set of physical
parameterizations appropriate for the Martian atmosphere as described in the 
works by \citet{Kur05,Kur13}. The MGCM accounts, among others, for radiative 
effects of gaseous carbon dioxide and airborne dust, and interactively
simulates condensation and sublimation of the atmospheric CO$_2$, formation of
CO$_{2}$ ice clouds, snowfalls and seasonal ice cap in the polar atmosphere.
The lower-resolution version of the MGCM has been validated against the 
observed zonal mean climatology \citep{Kur05}, and extensively been used for 
studies of baroclinic planetary waves \citep{Kur07}, zonal-mean variability 
in the middle- and high-latitudes \citep{Yam07}, equatorial semiannual 
oscillations \citep{Kur08}, winter polar warmings during global dust storms 
\citep{Kur09}, and CO$_2$ snowfalls in the northern winter polar atmosphere 
\citep{Kur13}.  Recently, this model received the name DRAMATIC (Dynamics, 
RAdiation, MAterial Transport and their mutual InteraCtions) MGCM, and has 
been used to validate the retrieved temperature in the southern polar night 
from the MGS radio occultation measurements \citep{Nog14}.

In this study, the MGCM was run at the T106 spectral truncation, which 
corresponds approximately to a 1.1$^{\circ}$ $\times$ 1.1$^{\circ}$ (or 
$\sim$60 km) horizontal resolution. In the vertical direction, the model 
domain extends from the surface to $\sim$80--100 km, and is represented by 
49 $\sigma$-levels.  Such setup allows for realistically capturing generation 
and propagation of GWs with horizontal wavelengths of $3\Delta x\sim$180 km 
and longer and, to some extent, their vertical attenuation due to nonlinear 
processes. These waves are subgrid-scale in conventional GCMs, and 
the dynamical and thermal importance in the Martian atmosphere of the 
harmonics of these scales has been demonstrated in the works of 
\citet{Med11b} and \citet{MY12}, correspondingly.

The local thermodynamic equilibrium was assumed for the radiative effects of
CO$_2$ gas at all heights.

\section{Gravity Wave Variations in the Lower (10-30 km) Atmosphere}
\label{sec:LowerAtm}

\begin{figure*}
\begin{center}
\includegraphics[width=0.38\textwidth,angle=270]{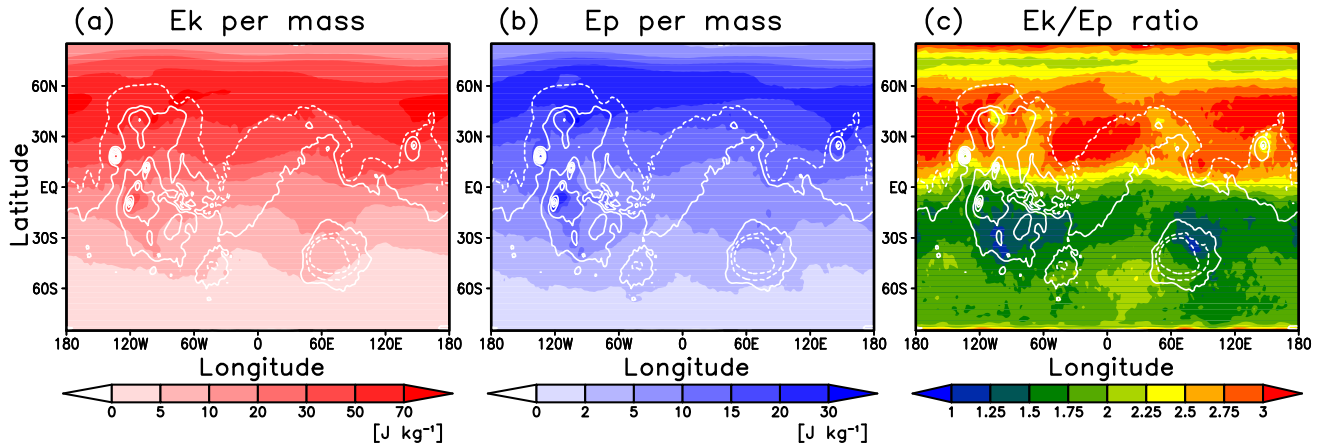}
\end{center}
\caption{(a) Kinetic $E_k$ and (b) potential energy $E_p$ per unit mass 
(in J~kg$^{-1}$) of resolved gravity waves with the total wavenumbers greater 
than 60 (horizontal wavelengths of less than $\sim$350 km), and (c) the ratio 
$E_k/E_p$, averaged between 10 and 100 Pa for 20 sols starting at 
$L_{s}$=270$^{\circ}$. White contours on each plot denote the Martian
topography.} 
\label{fig:1}
\end{figure*}

The results shown here are for the Northern Hemisphere winter solstice, i.~e., when Mars is
at perihelion, and the dynamical processes in the atmosphere are most active. All the
figures are based on 20-sol averaged fields centered at the solar longitude $L_s=270^\circ$,
with the dust opacity of $\sim$1.0 in the visible wavelength (a ``low dust" condition).

We designate the shortest horizontal-scale fluctuations with the total wavenumber $n>61$
(horizontal wavelengths less than $\sim$350 km) as wave disturbances $\varphi^\prime$. This
choice allows for explicitly considering harmonics, which are known to significantly
contribute to dynamical and thermal forcing of large-scale atmospheric flows, and which are
usually parameterized in GCMs. Correspondingly, the larger-scale ($n\le60$) fields here
represent the ``mean" $\bar{\varphi}$ such that
$\varphi=\bar{\varphi}+\varphi^\prime$. Similar definition is applicable to disturbance
covariances. For instance, $\overline{\varphi^\prime \psi^\prime}$ is the product of
shorter-scale fields $\varphi$ and $\psi$ on the globe, of which only the lower-$n$ portion
is taken. Effectively, averaging denoted by overbars is a horizontal spatial averaging, or a
coarse-graining.

Direct measures of activity of fluctuating fields, which we believe are composed mainly of
gravity waves, are their kinetic and potential energy (per unit mass) $E_k$ and $E_p$,
correspondingly:
\begin{equation}
  E_k={1\over 2} \biggl( \overline{u^{\prime 2}} + 
    \overline{v^{\prime 2}} \biggr),
\qquad
E_p={1\over 2} \Bigl(\frac{g}{N}\Bigr)^2 
    \frac{\overline{T^{\prime 2}}}{\overline{T}^2},
\label{eq:energy}
\end{equation}
where $u^\prime$ and $v^\prime$ are the wind fluctuations in the zonal and
meridional directions, respectively, $g$ is the acceleration of gravity, and 
$N$ is the Brunt-V\"ais\"al\"a frequency. The quantities $E_k$ and $E_p$ 
averaged between 10 and 30 km are shown in Figures~\ref{fig:1}a and 
\ref{fig:1}b. This representation allows for a direct comparison with the 
measurements of GW temperature fluctuations derived from MGS occultation data 
for the same season \citep[][Figure 4b]{Cre06a}. Their observations show a 
gradual increase of $E_p$ in the Southern Hemisphere from $<$2 J~kg$^{-1}$ at 
high-latitudes to 10--15 J~kg$^{-1}$ and larger over the equator, which is in 
an excellent agreement with our simulations in Figure~\ref{fig:1}b. 
Measurements are missing for latitudes higher than 20$^\circ$N, where 
simulations predict an increase of GW activity, and reaches its maximum (of 
greater than 30 J~kg$^{-1}$) over the core of the westerly polar night jet (at 
$\sim$60$^\circ$). Another observational constraint have been presented 
by \citet{Wri12}, who derived temperature fluctuations from the Mars Climate 
Sounder (MCS) data. Although they were obtained for spatial scales longer than 
in our simulations, the magnitudes of variations are in a very good agreement
between 100 and 10 Pa (several K) \citep[][Figure 2a]{Wri12}. Observations 
also show an enhancement of temperature fluctuations in the Northern 
high-latitudes. In the Southern Hemisphere, our simulations do not reproduce 
large temperature fluctuations. In addition, \citet{And12} showed that the
equatorial region has larger $E_p$ than any other latitude region, whereas
our simulations display that this peak is shifted to the middle latitudes of 
the Northern Hemisphere. 

The distribution of $E_k$ in Figure~\ref{fig:1}a is similar to that of $E_p$.  
It also demonstrates the latitudinal asymmetry of gravity wave activity in the 
lower atmosphere with the maximum in the winter hemisphere. There is an 
equipartition of kinetic energy between the zonal and meridional components of 
small-scale wind variations. The results in Figure~\ref{fig:1}a indicate that 
the magnitudes of wind fluctuations increase from $\sim$1 m~s$^{-1}$ in high 
latitudes of the Southern Hemisphere to $\sim$6 m~s$^{-1}$ in the middle- and 
high latitudes of the Northern Hemisphere. These distributions of $E_p$ and 
$E_k$ clearly reflect GW sources in the lower atmosphere.

One property of the small horizontal-scale wave field can immediately be 
found by comparing $E_k$ and $E_p$: the kinetic component of energy exceeds
that of potential energy. \citet[][Equation (10)]{geller10} have derived
the relation between the $E_k/E_p$ ratio and the intrinsic frequency of
gravity wave $\hat{\omega}$:
\begin{equation}
\frac{E_k}{E_p} = \frac{1+\biggl({f \over \hat{\omega}} \biggr)^2 }
                       {1-\biggl({f \over \hat{\omega}} \biggr)^2 },
\label{eq:ratio}
\end{equation}
where $f$ is the Coriolis frequency. It follows from (\ref{eq:ratio}) that
smaller-$\hat{\omega}$ (longer-period in the frame of reference moving with 
the local wind) GW harmonics have larger $E_k/E_p$ ratios, while the latter
asymptotically approaches unity for high-frequency harmonics. The calculated 
ratio $E_k/E_p$ plotted in Figure~\ref{fig:1} points out the 
interhemispheric asymmetry in the distribution of the dominant intrinsic 
frequencies $\hat{\omega}$ of resolved small-scale waves: they are 
a factor of two or more smaller in the winter 
hemisphere. Given that their horizontal scales are approximately equal 
throughout the globe, this implies smaller intrinsic horizontal phase 
velocities $c-\bar{u}$ of GWs in the Northern Hemisphere. These waves are 
generated by the meandering strong winter polar jet (large $\bar{u}$), which 
means that their observed horizontal phase velocities $c$ (measured with
respect to the surface) are, on the contrary, large. A closer consideration 
of small-scale GW-induced fields, for instance, of the horizontal wind 
divergence $\partial u^\prime/\partial x +\partial v^\prime/\partial y$ 
(see Movie S1 in the Supporting Information), confirms that wave packets move 
eastward much faster in the winter polar jet region, although somewhat lag 
the mean zonal winds. This illustrates the bias in the 
horizontal phase velocities of small-scale GWs in the source region first 
pointed out in the work by \citet{mkb98} and utilized in the prescribed source 
spectrum in the GW parameterization studies for Earth \citep{Yigit_etal09} 
and Mars \citep{Med11b}. In the mountainous regions, $E_k/E_p$ is, on the
contrary, small (blue shades in Figure~\ref{fig:1}c), which indicates large
intrinsic/small observed horizontal phase velocities. This means that
topographically-induced GWs dominate there, and that the wave packets are 
``tied up" to the relief features. Movie S1 clearly demonstrates this 
phenomenon.

\section{Vertical Propagation of Gravity Waves}
\label{sec:vert}

\begin{figure*}
\begin{center}
\includegraphics[width=0.7\textwidth,angle=270]{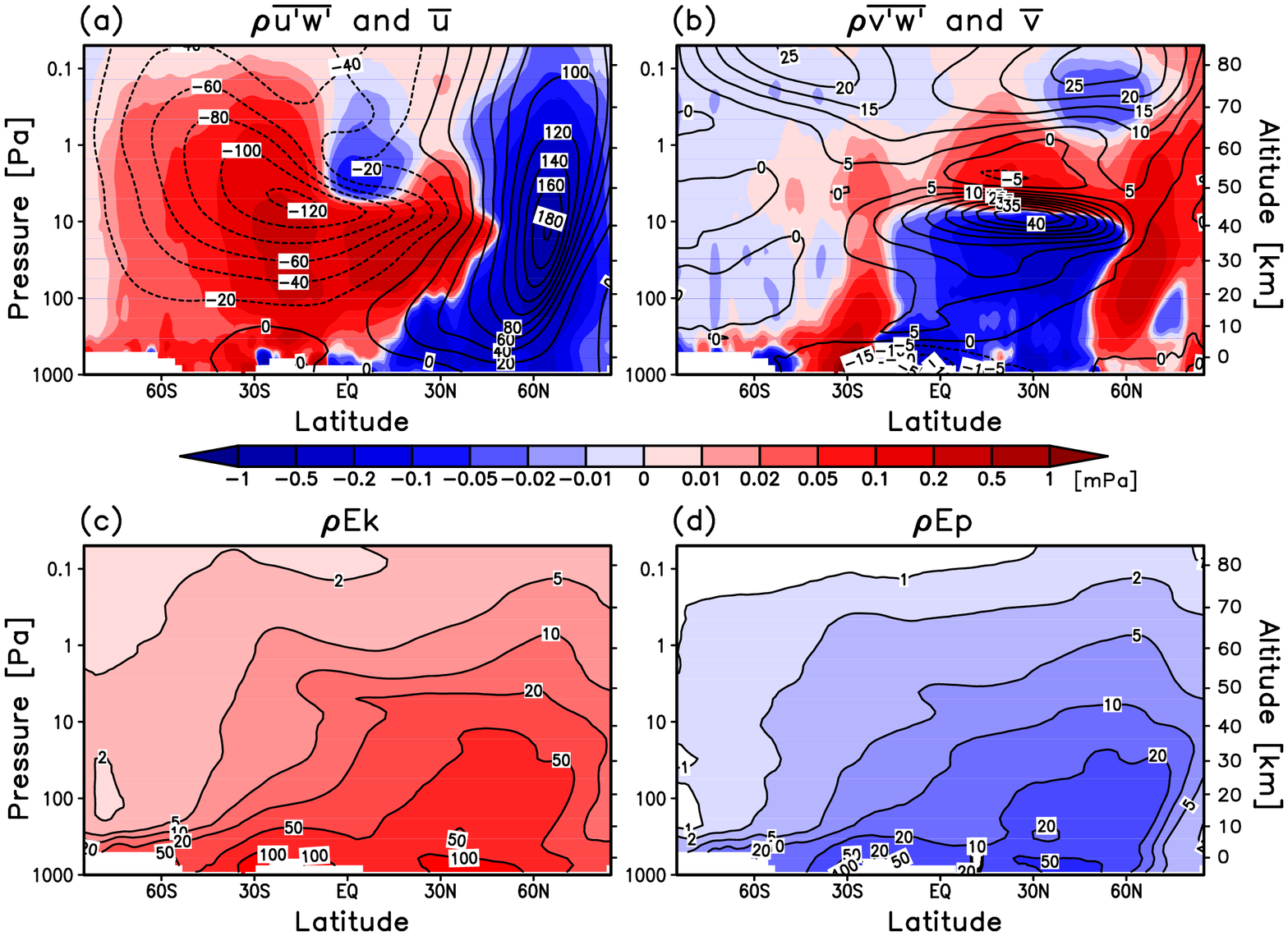}
\end{center}
\caption{The latitude-altitude cross-sections of zonal-mean quantities due 
to resolved GWs with the total wavenumber of larger than 60 (shaded): 
(a) vertical flux of zonal wave momentum $\rho\overline{u'w'}$ (in mPa), 
(b) vertical flux of meridional momentum $\rho\overline{v'w'}$, 
(c) kinetic wave energy $\rho E_k$ (in mJ~m$^{-3}$), 
(d) potential wave energy $\rho E_p$. 
Black contours in (a) represent the zonal wind (in m~s$^{-1}$), 
and the meridional wind in (b). }
\label{fig:2}
\end{figure*}

\begin{figure*}
\begin{center}
\includegraphics[width=0.7\textwidth,angle=270]{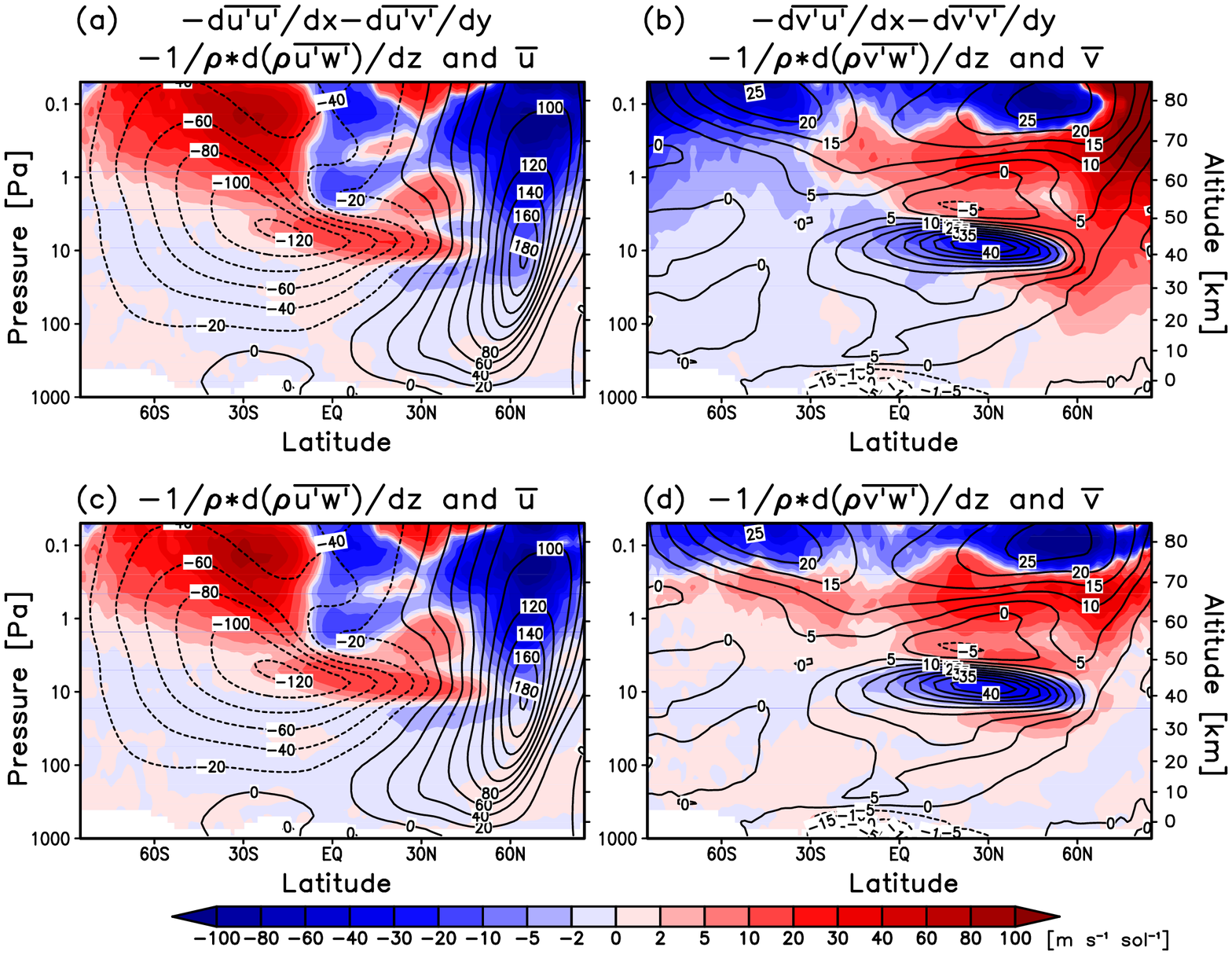}
\end{center}
\caption{The components of divergences of gravity wave momentum fluxes 
(shaded, in m~s$^{-1}$~sol$^{-1}$) and the mean wind (contours, in m~s$^{-1}$): full (horizontal and vertical)
divergences of (a) zonal and (b) meridional momentum fluxes; 
only vertical divergence of (c) zonal and (d) meridional momentum fluxes. 
Black contours denote (a, c) the mean zonal, and (b, d) meridional wind.}
\label{fig:3}
\end{figure*}

Having considered GWs in the lower atmosphere, we now turn to their upward
propagation. Vertical fluxes of the zonal and meridional momentum,
$\rho\overline{u^\prime w^\prime}$ and $\rho\overline{v^\prime w^\prime}$, respectively, are
important quantities for examining this. Zonally averaged distributions of the calculated
$\rho\overline{u^\prime w^\prime}$ and $\rho\overline{v^\prime w^\prime}$ are plotted with
color shades in Figures~\ref{fig:2}a and b, respectively, and the mean zonal and meridional
winds are superimposed with contour lines. The fluxes are vector quantities, which are
conserved if no sources and sinks are present. For a given GW harmonic, the momentum flux is
proportional to the intrinsic phase velocities in the corresponding direction, and
characterizes wave propagation with respect to the mean flow. Only in the absence of the
latter, the signs indicate the direction of wave propagation with respect to the surface,
that is, in the east-west or north-south. These results suggest that, in the lower
atmosphere, the fluxes are, generally, directed against the local winds. This means that the
spectra of GWs are dominated by harmonics with observed phase velocities $c$ that are slower
than the local wind (``lagging" the flow), or having opposite signs (moving against the
flow): $c<\bar{u}$ if $\bar{u}>0$, and $c>\bar{u}$ if $\bar{u}<0$.  Over the course of
vertical propagation, harmonics are selectively dissipated and/or obliterated due to
breaking or filtering by the mean wind. The net wave momentum flux is determined by a
delicate balance of contributions of ``surviving" harmonics from the initial spectrum. Thus,
magnitudes and even the sign of the net flux can vary with height. For instance, the
apparent increase of the magnitude in low latitudes between 100 and 10 Pa in
Figure~\ref{fig:2}a does not necessarily indicate in situ generation of waves with positive
$\rho\overline{u^\prime w^\prime}$.  Harmonics with $\rho\overline{u^\prime w^\prime}<0$ in
the incident spectrum are filtered by the easterly wind $\bar{u}<0$, while waves carrying
positive fluxes progressively contribute more, because a) their amplitudes grow with height,
and b) $c-\bar{u}$ and the associated momentum flux increase.  Above $\sim$10 Pa, the
opposite occurs. Harmonics with positive flux partly dissipate and deposit their momentum to
the mean flow, as we shall discuss below, and partly their contribution decreases (due to
the mean zonal wind $\bar{u}$ weakening) along with the increase of the contribution of
waves with negative fluxes.  Same can be applied to the local maximum of positive meridional
fluxes $\rho\overline{v^\prime w^\prime}>0$ over $\sim$60$^\circ$N in Figure~\ref{fig:2}b.

Since wave momentum fluxes are vector quantities they are not fully suitable for
characterizing the net field, because harmonics with opposite signs may offset and even
cancel contributions of each other. Wave variances provide another proxy for wave activity,
which is devoid of this limitation.  Figures~\ref{fig:2}c and d show their zonal mean
latitude-altitude distributions in the form of kinetic and potential energy, $E_k$ and $E_p$
from (\ref{eq:energy}), multiplied by the mean density. $\rho E_k$ exceeds $\rho E_p$
everywhere in the atmosphere, as it does at lower altitudes in Figure~\ref{fig:1}. The
maximum of wave energy is in the lower atmosphere, where these waves are mainly excited, and
decreases with height in each vertical column. However, a clear asymmetry between the
Northern and Southern Hemispheres is seen. Wave activity is stronger, and GWs penetrate
higher in the winter hemisphere. Partially, this may be explained by the asymmetry of
sources in the lower atmosphere, but refractive properties of the atmosphere associated with
the mean winds are likely to play a role as well. Spectra of generated waves in average, are
dominated by harmonics with slower phase velocities, as otherwise would cause an
``ultraviolet catastrophe" (integral of energy over spectrum diverges). These waves are less
affected by strong winds in the core of the westerly jet, and, therefore, are being focused
into it. One more reason for the asymmetry can be related to the oblique propagation: wave
packets composed of harmonics with slower phase velocities can cover significant horizontal
distances upon their vertical propagation.  We cannot diagnose the degree of obliqueness
directly from the GCM output, and a ray tracing model is required for that. Most likely, all
three factors contribute to the obtained distributions of GW activity in the middle
atmosphere. Here we simply state that the simulated asymmetry awaits a validation with
observations, and that any successful parameterization of subgrid-scale GWs must reproduce
it.

Figures~\ref{fig:2}a and b show that momentum fluxes ultimately decrease with
height. Divergence of the momentum fluxes quantifies the rate of wave 
obliteration, and the amount of momentum transferred to the mean (larger-scale
in our study) flow. Depending on the sign, waves can produce acceleration or
deceleration of the latter.
Figures~\ref{fig:3}a and b present thus calculated forcing along the 
corresponding axes:
  \begin{equation}
    a_x=-\nabla\cdot\overline{{\bf v'}u'}, \qquad
    a_y=-\nabla\cdot\overline{{\bf v'}v'},
    \label{eq:acc}
  \end{equation}
where ${\bf v'}=(u',v',w')$ are the components of velocity fluctuations, and
$\nabla =(\partial/\partial x, \partial/\partial y, \rho^{-1} \partial \rho /
\partial z)$.
As can be seen, $a_x$ and $a_y$ created by the resolved small-scale motions are significant
in the middle atmosphere (tens of m~s$^{-1}$~sol$^{-1}$), and directed mainly against the
mean wind. This result is consistent with the estimates of GW drag obtained using the
extended spectral parameterization of \citet{Yigit_etal08} applied to the distributions of
wind and temperature from the Mars Climate Database \citep{Medvedev_etal11a}, and
interactively coupled with the Martian GCM \citep{Med11b}. One may notice that it is
significantly smaller than the estimates of \citet{Fri06} ($\sim$1000 of m s$^{-1}$
sol$^{-1}$), but it is because we present zonal and time averaged quantities, while their
results are based on individual measurements. Instantaneously, $a_x$ and $a_y$ in our
simulations can reach over 10 000 m~s$^{-1}$~sol$^{-1}$. The response of the mean zonal
winds to this forcing is also seen -- the jets show the tendency to decrease and close in
the upper portion of the domain.  This cannot be achieved in simulations with conventional
(low) resolution without parameterized subgrid-scale GWs, unless an artificial sponge layer
is applied near the top. Thus, our GW-resolving simulations represent a direct confirmation
of the predictions on the dynamical importance and effects of small-scale GWs in the Martian
atmosphere.

The plotted divergences further illustrate GW propagation in the equatorial region. They
show weak negative $a_x$ below 10 Pa created by the absorption of harmonics with negative
fluxes by the easterly mean wind $\bar{u}<0$, as is discussed above. Around 10 Pa, strong
dissipation of harmonics with $\overline{u^\prime w^\prime}>0$ produces positive $a_x$
decelerating the mean wind. Above 10 Pa, the remaining harmonics with negative fluxes
deposit the negative momentum upon their dissipation, which results in the acceleration of
the negative flow. The latter seems paradoxial as all the waves with negative fluxes should
have apparently been filtered below by the negative background wind. An in depth explanation
of such phenomenon was given in the paper of \citet[][Section 8, paragraph 42 and Figure
8]{Yigit_etal09}, and is related to the fact that the projection of the wind on the
direction of wave propagation (that affects the latter) can significantly differ from the
zonal wind alone.
  
The vertically alternating patches of positive and negative $a_x$ in the equatorial region
tend to enhance the semiannual oscillation of the zonal wind, as discussed in the work of
\citet{Kur08}. The meridional component of the GW-induced torque, $a_y$, also plays an
important role in the middle atmosphere. It decelerates the cross-equatorial south-to-north
meridional transport in low and middle latitudes at $\sim$10 Pa induced mainly by thermal
tides, accelerates it somewhat higher (at $\sim$1 Pa), and extends to high latitudes of the
winter (Northern) hemisphere. This leads to the intensification of the downward branch of
the meridional transport cell over the North Pole, which results in the increase of the
adiabatic heating and enhancement of the middle atmosphere polar warming \citep{MH07,Kur09}.
Similarly, small-scale GWs decelerate the northward meridional flow in the upper mesosphere,
and weaken the the meridional pole-to-pole cell.

Next, we estimate the contributions of the vertical component of the momentum flux
divergence to the net $a_x$ and $a_y$ by plotting
$-\rho^{-1}d\rho \overline{u^\prime w^\prime}/dz$ and
$-\rho^{-1}d\rho\overline{v^\prime w^\prime}/dz$ in Figures~\ref{fig:3}c and d.  They are
very close to those in Figures~\ref{fig:3}a and b. This indicates that a) horizontal
propagation of GWs plays a secondary role in forcing the mean flow, and b) GW
parameterizations accounting for only vertical propagation can successfully capture the
major part of subgrid-scale GW effects in GCMs.

\section{Horizontal Distributions of Wave Fluxes}
\label{sec:horiz}

\begin{figure*}
\begin{center}
\includegraphics[width=0.7\textwidth,angle=270]{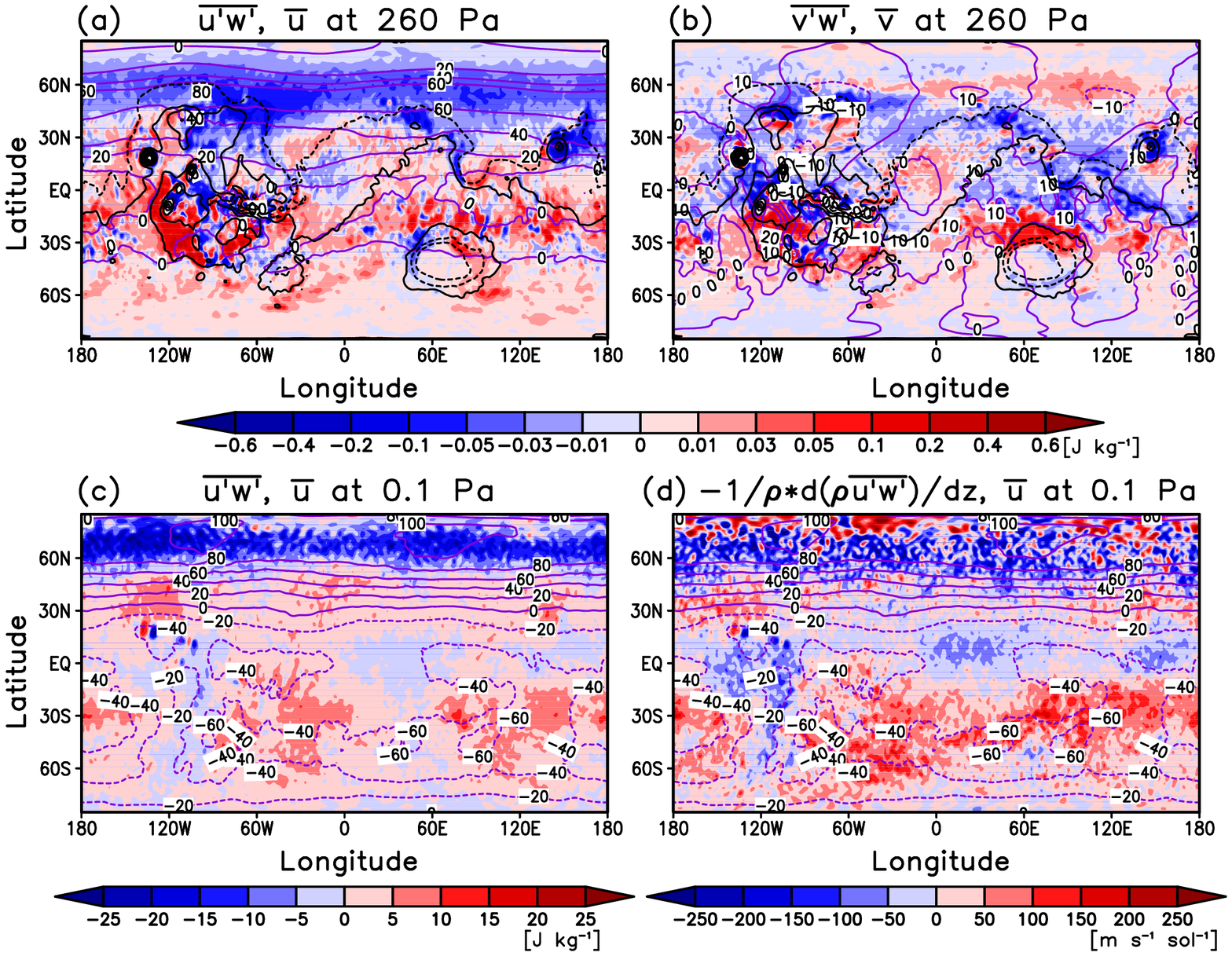}
\end{center}
\caption{The distribution of horizontal flux of zonal wave momentum $\overline{u'w'}$ at (a)
  260 Pa and (c) 0.1 Pa pressure levels (shaded, in J~kg$^{-1}$).  Black contours represent
  the topography of Mars. (b) The same as in (a) except for $\overline{v'w'}$. (d) The
  divergence of meridional wave momentum (in m~s$^{-1}$~sol$^{-1}$) at 0.1 Pa pressure
  level. Black contours in (a) and (b) represent the topographical features. Purple contours
  in (a), (c) and (d) denote the mean zonal wind velocity $\bar{u}$ (in m~s$^{-1}$), while
  in (b) denote the mean meridional wind velocity $\bar{v}$.}
\label{fig:4}
\end{figure*}

Many parameterizations use wave momentum fluxes at a certain level in the lower atmosphere
for the specification of sources. Therefore, we plotted the longitude-latitude distributions
of $\overline{u^\prime w^\prime}$ and $\overline{v^\prime w^\prime}$ at $p=260$ Pa in
Figures~\ref{fig:4}a--b.  They are shown with color shades, and the corresponding
large-scale winds $\bar{u}$ and $\bar{v}$ are superimposed with contours. Although all
quantities are 20-day averaged, fluxes are seen to be very patchy, which demonstrates that
sources are extremely localized both in space and time.  Peak values of the fluxes with
alternating signs occur in the mountainous regions. They are evidently associated with waves
generated by flow over topography. Nevertheless, a clear asymmetry can be seen: fluxes
predominantly have signs opposite to the mean local wind. In the middle- to high-latitudes
of the Northern Hemisphere, the distribution of the zonal flux is significantly
smoother. These GWs are excited within the curvatures of the winter westerly jet, which, in
large, are associated with Kelvin waves moving eastward with time. The meridional fluxes are
negative and directed against the mean meridional wind between the Equator and 45$^\circ$N,
and have alternating direction in other regions, where the mean wind is weak.

For comparison, the zonal momentum fluxes created by harmonics penetrating to 
the mesosphere ($p=0.1$ Pa) are shown in Figure~\ref{fig:4}c. 
Their distribution is significantly more horizontally homogeneous. 
Most orographic GWs (with small with respect to the surface phase 
speeds) are filtered out by the wind in the course of their vertical 
propagation, and create only a marginal enhancement over the mountainous
regions. This confirms the fact well-known from Earth studies that GWs with
progressively faster horizontal phase speeds dominate at high altitudes
\citep{YigitMedvedev15}. The region with negative (but large) 
horizontal wave momentum fluxes in the mesosphere is confined to Northern
high-latitudes, which reflects the favorable propagation conditions for the 
corresponding harmonics, and which is in line with our finding using the GW 
parameterization \citep{Med11b, Yigit_etal15}. The magnitudes of fluxes in 
the mesosphere significantly exceed those in the lower atmosphere, which 
merely reflects the wave amplitude growth due to exponential density drop 
with height.

Finally, we show the calculated vertical divergence of momentum fluxes (wave
drag), $a_x=-\rho^{-1} d(\rho\overline{u^\prime w^\prime})/dz$, in the 
mesosphere (Figure~\ref{fig:4}d). It is consistent with the zonal mean 
cross-section in Figure~\ref{fig:2}d, but shows a high degree of horizontal
inhomogeneity. Locally, $a_x$ exceeds 200 m~s$^{-1}$~sol$^{-1}$ at $p=0.1$ Pa,
but almost nowhere is less than several tens of m~s$^{-1}$~sol$^{-1}$.
Obviously, such strong effects of small-scale waves cannot be ignored in the
dynamics of the Martian mesosphere. Note that the values and distributions of
both wave fluxes and acceleration/deceleration obtained in this high-resolution
simulation can and should be served for validation and tuning of GW
parameterizations.

\section{Conclusions}
\label{sec:concl}

We presented the first results of simulations with a new high-resolution 
Martian general circulation model (triangle spectral truncation T106) that 
resolves (in a 3$\Delta x$ sense) harmonics with horizontal scales down 
to $\sim$180 km. In this paper, we concentrated on the Northern winter 
solstice (around the solar longitude $L_s=270^\circ$) and GW harmonics 
shorter than 350 km. This consideration leaves aside shorter-scale (few tens 
of km) harmonics generated by convection, and which can be important in the 
upper atmosphere. The main inferences of this first study of its kind are 
listed below.

\begin{enumerate}
\item Magnitudes of temperature variances due to small-scale GWs (or available
  potential wave energy $E_p$) between 10 and 30 km are in a good agreement 
  with those obtained by \citet{Cre06a} from Mars Global Surveyor radio 
  occultation data. In addition, simulations show a gradual and steep 
  latitudinal increase of $E_p$ from South to North with the maximum in the 
  winter hemisphere, where the observational data are missing.

\item Variances of wave-induced horizontal wind fluctuations exhibit a similar 
  behavior, however, with a steeper growth -- the ratio of the wave kinetic 
  and potential energy, $E_k/E_p$, increases from $\sim$1.5 in the Southern 
  Hemisphere to about 3 in the Northern one.

\item Two major sources of GWs can be identified in the lower atmosphere: the
  mountainous regions generating slow, or even non-moving with respect to
  the surface wave packets, and the meandering winter westerly jet exciting
  faster GW harmonics traveling mainly eastward.

\item The majority of generated GWs move slower than the background 
  wind, and the associated vertical fluxes of horizontal wave momentum 
  are directed against it.

\item Most of GWs are produced in the lower atmosphere, and their fluxes and
  energy decay with height.

\item Upon vertical propagation and dissipation, these waves deposit their
  momentum directed mainly against the local wind, and, thus, provide a wave
  drag on the mean flow.

\item As a result of the drag, the simulated jets in both hemispheres 
  demonstrate a tendency to close in the upper atmosphere. This feature 
  cannot be reproduced by GCMs with a conventional (low) resolution without 
  applying an artificial sponge near the model top or an appropriate GW
  parameterization.

\item In the lower atmosphere, the distributions of wave momentum fluxes
  are very patchy, reflecting the highly localized nature of GW sources. 
  Orographically generated slow waves are filtered in lower layers in the 
  course of their vertical propagation, and the upper mesosphere is dominated 
  by harmonics with faster horizontal phase velocities.
\end{enumerate}

Given the lack of observations of GWs in the atmosphere of Mars, our
high-resolution simulations provide the much needed framework for constraining 
GW parameterizations, and validating the results obtained with the latter.

\acknowledgments{
Data supporting the figures are available from TK 
(tkuroda@\-pat.\-gp.\-tohoku.\-ac.jp).

TK was supported by the Japan Society for the Promotion of Science (JSPS)
KAKENHI Grant Number 24740317, and the Promotion of the Strategic Research
Program for Overseas Assignment of Young Scientists and International
Collaborations titled ''Intensification of International Collaborations for
Planetary Plasma and Atmospheric Dynamics Research based on the Hawaiian
Planetary Telescopes''. The model runs have been performed with the HITACHI
SR16000 System (yayoi) at the Information Technology Center, The University of
Tokyo. This work was partially supported by German Science Foundation (Deutsche
Forschungsgemeinschaft) grant ME2752/3-1 and NASA grant NNX13AO36G.
}

\end{document}